\documentclass[manuscript]{aastex}
\usepackage{aas_macros,url,natbib,amssymb,amsmath,CJK,lineno}

\shorttitle{Comet 252P/LINEAR: born (almost) dead?}
\shortauthors{Ye et al.}

\begin{document}
\begin{CJK*}{UTF8}{gbsn}

\title{Comet 252P/LINEAR: born (almost) dead?}

\author{Quan-Zhi Ye (叶泉志)}
\affil{Department of Physics and Astronomy, The University of Western Ontario, London, Ontario N6A 3K7, Canada}
\email{qye22@uwo.ca}

\and

\author{Peter G. Brown and Paul A. Wiegert}
\affil{Department of Physics and Astronomy, The University of Western Ontario, London, Ontario N6A 3K7, Canada}
\affil{Centre for Planetary Science and Exploration, The University of Western Ontario, London, Ontario N6A 5B8, Canada}

\begin{abstract}
Previous studies have revealed Jupiter-family comet 252P/LINEAR as a comet that was recently transported into the near-Earth object (NEO) region in $\sim1800$~AD yet only being weakly active. In this Letter, we examine the ``formed (almost) dead'' hypothesis for 252P/LINEAR using both dynamical and observational approaches. By statistically examining the dynamical evolution of 252P/LINEAR over a period of $10^7$~years, we find the median elapsed residency in the NEO region to be $4\times10^2$~years which highlights the likelihood of 252P/LINEAR as an (almost) first-time NEO. With available cometary and meteor observations, we find the dust production rate of 252P/LINEAR to be at the order of $10^6$~kg per orbit since its entry to the NEO region. These two lines of evidence support the hypothesis that the comet was likely to have formed in a volatile-poor environment. Cometary and meteor observations during the comet's unprecedented close approach to the Earth around 2016 Mar. 21 would be useful for the understanding of the surface and evolutionary properties of this unique comet.
\end{abstract}


\section{Introduction}

Comets are small, icy objects originating from the outer solar system. They are the leftover planetesimals from the formation of the outer planets. These objects remain in the outer solar system for most of their lifetime, until perturbations sent them into the inner solar system where they become visible. It has been noted that the observable comet population displays a large diversity in ice composition which, apart from evolutionary effects, is linked to predictions that comet nuclei were formed at different places and times in the solar nebula \citep[e.g.][]{Whipple1987,Bockelee-Morvan2004}. Thus, the observation of comets provides a unique opportunity for understanding the range of chemistry in the primitive solar nebula.

However, the quest is not without its obstacles. Observational interpretations are biased by the fact that more active comets tend to be easier to detect and study, meaning that less active comets are somewhat underrepresented in the sample. Dynamical investigations are limited by the chaotic nature of the orbital evolution of small bodies, which make it challenging to reconstruct the orbit of history over even modest timescales (e.g. $\sim10^3$~yr). The fact that the evolutionary processes of cometary nuclei are little understood makes it difficult to isolate evolutionary effects from formation diversity when addressing the volatile inventory of individual objects.

Numerical integrations carried out by \citet{Tancredi2014h} indicate that Jupiter-family comet (JFC) 252P/LINEAR might have entered the near-Earth object (NEO) region\footnote{NEO region refers to the region within 1.3~AU from the Sun.} only $\sim50$ orbits ago. This short dynamical timescale suggests that 252P/LINEAR should be a ``physically young'' comet, considering the typical physical lifetime of near-Earth JFCs of 150--200 revolutions \citep{DiSisto2009}. However, with an absolute total magnitude of $M_1=18.6$\footnote{From JPL Small-Body Database, \url{http://ssd.jpl.nasa.gov/sbdb.cgi?sstr=252P}, accessed 2015 Dec. 21.}, 252P/LINEAR exhibits the characteristics of a weakly active comet \citep{Ye2016}. Considering its recent entry to the NEO region, we expect evolutionary processes to have played a relatively minor role in altering the subsurface volatile composition. One possible explanation of the weak activity in this young comet could be a volatile-poor environment at the time of formation of the nucleus.

The 2016 perihelion passage of 252P/LINEAR offers an exceptional opportunity for Earth-based observers to study the comet. The comet will pass perihelion on 2016 Mar. 15 and make a close approach to the Earth on 2016 Mar. 21 at 0.036~AU, which is one of the closest cometary approaches to the Earth on record\footnote{\url{http://www.minorplanetcenter.net/iau/lists/ClosestComets.html}, accessed 2015 Dec. 21.}. The Earth may also have passed through the dust trails produced by 252P/LINEAR during its past revolutions\footnote{M. Maslov (unpublished), \url{http://feraj.narod.ru/Radiants/Predictions/252p-ids2016eng.html}, accessed 2015 Dec. 21.}, potentially producing meteor activity. Despite the proximity of 252P/LINEAR's nodal point to the Earth's orbit, meteor activity from 252P/LINEAR has not been reported. However, it is well known that meteor observations are useful in understanding of the physical history of a comet \citep[e.g.][]{Yeomans1981, Jenniskens2004}. Meteoroids from 252P/LINEAR could provide clues to the recent history of the comet.

In this Letter, we examine the ``born (almost) dead'' scenario for the case of 252P/LINEAR in preparation for the observation campaigns in Mar. 2016. We adopt two approaches: an examination of the dynamical evolution of the comet (\S~2), as well as an analysis of the available comet and meteor observations guided by an existing cometary dust model (\S~3). Results from both tracks are discussed and summarized in \S~4.

\section{Dynamical Evolution}

We generated 1000 ``clones'' of 252P/LINEAR using its orbital covariance matrix provided in JPL~28\footnote{From JPL Small-Body Database, \url{http://ssd.jpl.nasa.gov/sbdb.cgi?sstr=252P}, accessed 2015 Dec. 21.}, and integrated them $10^5$~yr backwards in time from 2000~AD. Integration was performed with the MERCURY6 package using the symplectic integrator \citep{Chambers1999a}. Clones are considered to have been ejected from the solar system when they reach a heliocentric distance of 100~AU.

\begin{figure*}
\includegraphics[width=\textwidth]{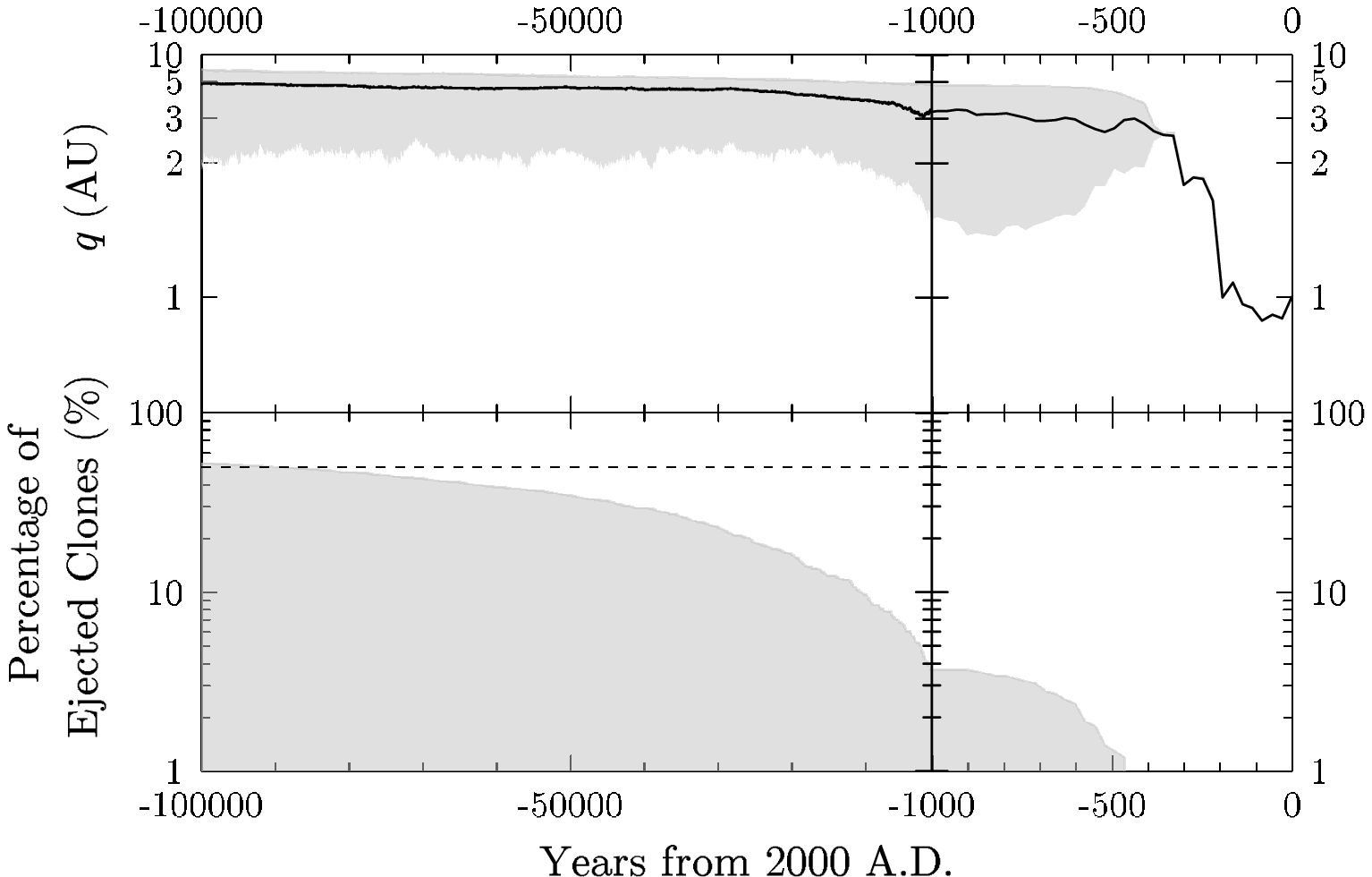}
\caption{Dynamical evolution of 1000 clones of 252P/LINEAR over a time interval of $10^5$~yr with a zoomed section for 1000~yr. Upper figure: median (black line) and $\pm1\sigma$ region (shaded) of the evolution of perihelion distance $q$. Lower figure: percentage of ejected clones (clones that reach heliocentric distance of 100~AU).}
\label{fig:clone}
\end{figure*}

As shown in Figure~\ref{fig:clone}, we confirm the recent (circa. 1800~AD) entry of 252P/LINEAR into the NEO region. The distribution of the clones is extremely compact until a close approach to Jupiter in 1690~AD (miss distance $\sim0.1$~AU). There have been a few recent approaches to the Earth: 0.10~AU in Mar. 2000 (when the comet was discovered), 0.08~AU in Mar. 1921, and 0.05~AU in Feb. 1847. If the activity level of 252P/LINEAR in the 19th century is comparable to what it is today, the comet would have been at +10 mag during its approach in 1847. We searched through the catalog of 19th century comets compiled by \citet{Kronk2008} but do not find any association. Comets at this magnitude may be at the observational limit for 19th century observers \citep{Everhart1967}, but the lack of detection also implies that the comet was probably not substantially much more active at that time compared to today.

With evidence that 252P/LINEAR has possibly been very weakly active since its current entry to the NEO region, it is possible that the comet depleted most of its volatiles during a previous residency in the NEO region. The question then becomes: how long did 252P/LINEAR reside in the NEO region in the past? To answer this question, we extend our integration backward and forward in time for $10^7$~years. We find the median elapsed residency in the solar system (i.e. the time that the clone \textit{already} spent as a bounded object) to be $1.3\times10^5$~years and median elapsed NEO region residency to be $4\times10^2$~years. The median dynamical lifetime (i.e. the \textit{total} time that the clone has and will spend as a bounded object) and median integrated time as a NEO are found to be $2.5\times10^5$~years and $4\times10^3$~years respectively. While this dynamical and NEO lifetime is comparable to the values for the general JFC population \citep{Fernandez2002n}, we note that the elapsed residency as a NEO is virtually null as it is certain the comet has already spent $\sim200$~years in the NEO region (since its current entry in $\sim1800$~AD). This suggests that 252P/LINEAR is likely a first time (or, almost first time) visitor to the NEO region.

We also examine the time that 252P/LINEAR has spent in a ``visible'' state \citep[i.e. $q<2.5$~AU,][]{Quinn1990}. The value of 2.5~AU is considered as the distance from the Sun that water ice sublimation is expected to start \citep[e.g.][]{McNamara2004l}. We find the elapsed and total median time of 252P/LINEAR as a visible comet to be $4\times10^3$~yr and $1.2\times10^4$~yr, the latter of which is more in line to literature values \citep[$\sim8.5\times10^3$~yr as given by][]{Levison1994b} compared to that in the NEO region. This indicates that 252P/LINEAR may have been active for some time before it entered the NEO region; but the process of thermal devolatilization occurs on longer timescales for comets with larger $q$ \citep[e.g.][]{Jewitt2004}.
 
\section{Current and Past Dust Production}

The current and past dust production of comets can be constrained by cometary and meteor observations. A parameter commonly used as a proxy for cometary dust production is the $Af\rho$ parameter \citep{Ahearn1984}:

\begin{equation}
Af\rho = 4 r_\mathrm{H}^2 \varDelta^2 \rho^{-1} \cdot 10^{0.4(m_\odot-m)}
\label{eq:afrho}
\end{equation}

\noindent where $r_\mathrm{H}$ is the heliocentric distance of the comet, $\varDelta$ is the geocentric distance of the comet, $\rho$ is the linear radius of the photometric aperture at the distance of the comet, and $m_\odot$ and $m$ are the apparent magnitudes of the Sun and the comet.

Prior to its 2016 perihelion passage, 252P/LINEAR had only been observed during its 2000 and 2010 passages (the 2005 passage was not observed possibly due to poor viewing geometry). However, the 2010 passage was only observed at $r_\mathrm{H}=2.5$~AU on its outbound leg and the comet had likely already ceased activity. Hence, we only use the observations from the 2000 passage which cover a section of $r_\mathrm{H}=1.1$~AU to 1.5~AU on the outbound leg of the comet. We obtain a total of 80 observations from the Minor Planet Center Observation Database (\url{http://www.minorplanetcenter.net/db_search/show_object?object_id=252P}) and calculate the $Af\rho$ quantity using Equation~\ref{eq:afrho}. The $\rho$ parameter accounts for the photometric aperture which differs among observers; to simplify the problem, we adopt an aperture of $13''$ from the conservative (small) end of coma diameter estimations \citep{Shelly2000a}. The computed $Af\rho$ are therefore at the upper end of possible values.

\begin{figure}
\includegraphics[width=0.5\textwidth]{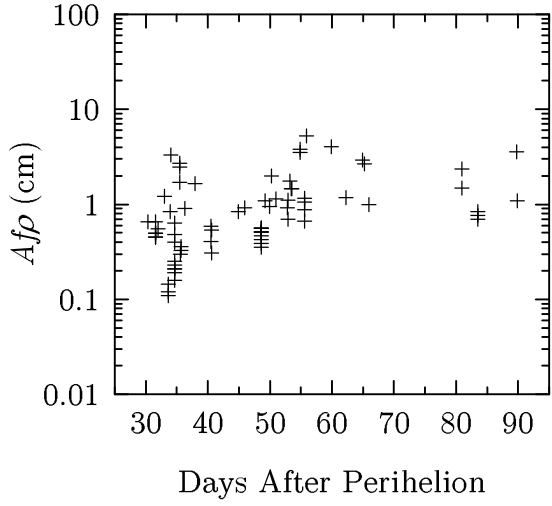}
\caption{Temporal variation of the $Af\rho$ parameter of 252P/LINEAR during its 2000 perihelion passage. The median value is 0.6~cm.}
\label{fig:afrho}
\end{figure}

The temporal variation of the $Af\rho$ parameter of 252P/LINEAR during its 2000 perihelion passage is shown as Figure~\ref{fig:afrho}. The upward trend before $t_\mathrm{P}+55$~d ($t_\mathrm{P}$ is the perihelion epoch) is likely an artifact due to small $\varDelta$ at the time of the observation and the conservative $\rho$ used in the calculation, resulting in an underestimation of the cometary flux \citep[coma diameter estimation given in][varies from $13''$ to $1'$]{Shelly2000a}. The median $Af\rho$ value is 0.6~cm, one of the lowest numbers ever measured for a comet and comparable with the reported $Af\rho$ of another low activity comet, 209P/LINEAR \citep{Schleicher2014c}. We further note that the $Af\rho$ may be contaminated by the light reflected from the nucleus given the extremely low dust production of 252P/LINEAR, which may result in some overestimation of the measured $Af\rho$; hence, the actual $Af\rho$ could be even lower. 

To search for any meteor activity originated from 252P/LINEAR, we make use of the Monte Carlo meteoroid stream model developed in our earlier works \citep[c.f.][and references therein]{Ye2015l}. Integrations are performed using the 15th order RADAU integrator bundled with the MERCURY6 package \citep{Everhart1985}. Gravitational perturbations from the eight major planets (with the Earth-Moon system represented by a single mass at the barycenter of the two bodies), radiation pressure, and Poynting-Robertson effect are included. We first integrate 252P/LINEAR back to 1800~AD (i.e. back to its entry into the NEO region) and then integrate it forward in time, releasing meteoroids in the diameter range of $0.5~\mathrm{mm}<a<50~\mathrm{mm}$ (i.e. the diameter range that associates with visible meteors) when the comet is at each perihelion, assuming a bulk density of $1~000~\mathrm{kg \cdot m^{-3}}$ and a size distribution of $\mathrm{d}N/\mathrm{d}a\propto a^{-2.6}$. Modeling of meteoroid streams is usually insensitive to the ejection model \citep[e.g.][]{Williams2001h, Williams2011a}; however, since 252P/LINEAR is apparently a low activity comet, we use two different ejection models and repeat the simulation for each one of them: the ``low activity scenario'' described in \citet{Ye2016} and the ``normal scenario'' described in \citet{Brown1998d}. At the completion of the integration, we examine two scenarios of possible meteor activity: (1) meteor outbursts from young meteoroid trails. This is examined by selecting meteoroids that are approaching the Earth within 0.01~AU in the period of interest; and (2) annual ``background'' activity, i.e. activity from older, more dispersed trails, this is examined by selecting meteoroids with Minimum Orbit Intersection Distance (MOID) with respect to the Earth's orbit $<0.01$~AU.

The simulation result for 2001--2020 is listed in Table~\ref{tbl:met}. We note that the result is sensitive to the choice of ejection model in contrast to other meteor showers, probably due to the relatively young age of the trails and small encounter speed. By applying the flux estimation technique outlined in \citet{Ye2015l}, we find that most of these computed activities are below the detection limit of modern meteor surveys, except for the outburst cases of 2011 and 2016. In 2011 the Earth passed close to the 1984- and 1989-trail, which are very young and compact trails formed recently. However, the characteristic Minimum Orbit Intersection Distance (MOID) of both encounters are close to the selection limit (close to $\sim0.01$~AU), therefore the possibility of meteor activity is likely minimal. In 2016, the Earth will pass close to the 1915-, 1921- and 1926-trails. Assuming the dust production rate of 252P/LINEAR in the 1920s is comparable to its current value, the meteor flux is likely to be at the order of $10^{-4}$ to $10^{-3}~\mathrm{km^{-2} \cdot hr^{-1}}$, or less than a few meteors in terms of Zenith Hourly Rate (ZHR).

\begin{figure}
\includegraphics[width=0.5\textwidth]{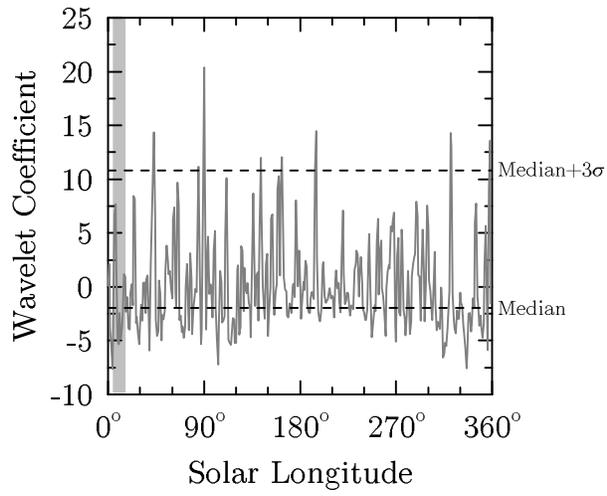}
\caption{Temporal variation of wavelet coefficient of CMOR data, centered at the computed radiant ($\alpha_\mathrm{g}=77^\circ$, $\delta_\mathrm{g}=-16^\circ$, $v_\mathrm{g}=10.9~\mathrm{km \cdot s^{-1}}$) of meteor activity from 252P/LINEAR. Shaded region indicates the predicted meteor activity from 252P/LINEAR.}
\label{fig:wc}
\end{figure}

Next, we conducted a search in the survey data collected by the Canadian Meteor Orbit Radar (CMOR) \citep[c.f.][and references therein]{Jones2005a}. We apply the wavelet technique described in \citet{Brown2008a} at the computed radiants to search for meteor activity, without finding any enhancement (Figure~\ref{fig:wc}). Using the number of background meteors detected by CMOR within a radius of $10^\circ$ of the expected radiant and $10\%$ from the expected meteoroid speed, as well as the CMOR collection area as a function of the declination \citep{Brown1995f}, we estimate the upper limit of the meteoroid flux to be at the order of $10^{-4}~\mathrm{km^{-2} \cdot hr^{-1}}$ to a limiting mass of $\sim10^{-7}$~kg, with an uncertainty within a factor of several due to the unconstrained mass distribution of the stream (which affects the collection area).

From the simulation, we find the meteoroid delivery efficiency (i.e. the fraction of the ejected meteoroids with MOID$<0.01$~AU) to be $\eta=17\%$. Assuming that the meteoroids are uniformly distributed along the orbit of 252P/LINEAR, we can estimate the past dust production since the comet's entry into the NEO region using 

\begin{math}
N = \frac{\displaystyle \mathcal{F} P \cdot \Delta X^2}{\displaystyle \eta N_\mathrm{orb}}
\end{math}

\noindent where $N$ is the meteoroid production per orbit, $\mathcal{F}=10^{-4}~\mathrm{km^{-2} \cdot hr^{-1}}$ is the upper limit of meteoroid flux constrained by meteor data, $P=5$~years is the orbital period of the meteoroids, $\Delta X=0.01$~AU is the collection area, and $N_\mathrm{orb}=50$ is the number of orbits that the comet is active. By inserting numbers from previous analysis, we get $N=10^{12}$ meteoroids, or $\sim10^6$~kg per orbit appropriate to millimeter-sized dust (assuming a bulk density of $1~000~\mathrm{kg \cdot m^{-3}}$). For comparison, the dust production rate of 55P/Tempel-Tuttle (parent of the Leonid meteor shower) is of the order of $10^{11}$~kg per orbit \citep{Vaubaillon2005b}. While the actual number may be off by an order of magnitude, such extremely low number illustrate 252P/LINEAR as a comet with very low dust production, as well as its lack of significant activity since its entry to the NEO region.

\section{Summary}

The two lines of evidence -- dynamical and observational -- have outlined 252P/LINEAR as a comet that is likely an (almost) first-time visitor to the NEO region, yet only little active in terms of dust production. These evidences support the hypothesis that 252P/LINEAR was likely to have formed in a volatile-poor environment, as compared to other members in the visible JFC population. Cometary and meteor observations during its close approach in Mar. 2016 will likely provide more information regarding the surface and evolutionary properties of this unique comet.

\acknowledgments

We thank an anonymous referee for his/her comments, as well as Michal Drahus and Davide Farnocchia for discussions. Thanks to Zbigniew Krzeminski, Jason Gill, Robert Weryk and Daniel Wong for helping with CMOR operations. Funding support from the NASA Meteoroid Environment Office (cooperative agreement NNX11AB76A) for CMOR operations is gratefully acknowledged.


\clearpage

\clearpage

\clearpage

\clearpage

\begin{table*}
\centering
\caption{Orbital elements and relevant parameters of 252P/LINEAR. Orbital elements are extracted from JPL~28.}
\begin{tabular}{ll}
\hline
Parameter & Value \\
\hline
Epoch & Julian Date 2455731.5 \\
Perihelion distance $q$ & 1.00006~AU \\
Semimajor axis $a$ & 3.05560~AU \\
Eccentricity $e$ & 0.67271 \\
Inclination $i$ & $10.38990^\circ$ \\
Longitude of the ascending node $\Omega$ & $190.99780^\circ$ \\
Argument of perihelion $\omega$ & $343.28753^\circ$ \\
Epoch of perihelion passage $t_\mathrm{p}$ & 2010 Nov. 13.66041 UT \\
Non-gravitational radial acceleration $\mathcal{A}_{1}$ & $3.25\times10^{-9}~\mathrm{AU \cdot d^{-2}}$ \\
Non-gravitational transverse acceleration $\mathcal{A}_{2}$ & $2.04\times10^{-10}~\mathrm{AU \cdot d^{-2}}$ \\
Non-gravitational normal acceleration parameter $\mathcal{A}_{3}$ & $3.36\times10^{-10}~\mathrm{AU \cdot d^{-2}}$ \\
Nucleus radius $R_\mathrm{N}$ & 0.5~km\tablenotemark{a} \\
\hline
\end{tabular}
\tablenotetext{a}{Drahus (2015, personal communication).}
\label{tbl:orb}
\end{table*}

\clearpage

\begin{table*}
\centering
\caption{Computed meteor activity from 252P/LINEAR in 2001--2020, with the low-activity ejection model from \citet{Ye2016} and normal ejection model from \citet{Brown1998d}.}
\begin{tabular}{llllcc}
\hline
Year & Ejection model & Trail & Peak Time & Radiant & $v_\mathrm{g}$ \\
     &                &       & (UT) & $\alpha_\mathrm{g}, \delta_\mathrm{g}$ & $\mathrm{km \cdot s^{-1}}$ \\
\hline
2002 & Low-activity & 1910 & 2002 Mar. 28 03:40 & $89.9^\circ, -12.4^\circ$ & 10.3 \\
     & Normal & 1910 & 2002 Mar. 28 06:23 & $90.0^\circ, -12.1^\circ$ & 10.3 \\
\hline
2008 & Low-activity & 1905 & 2008 Apr. 15 16:47 & $77.4^\circ, +4.6^\circ$ & 10.9 \\
     & Normal & 1905 & 2008 Apr. 15 18:50 & $77.5^\circ, +4.7^\circ$ & 10.9 \\
\hline
2011 & Normal & 1984 & 2011 Apr. 4 16:45 & $75.7^\circ, -9.1^\circ$ & 11.1 \\
     &        & 1989 & 2011 Mar. 31 2:11 & $77.4^\circ, -16.0^\circ$ & 11.0 \\
\hline
2016 & Low-activity & 1915 & 2016 Mar. 28 00:10 & $78.2^\circ, -17.7^\circ$ & 11.0 \\
     &              & 1921 & 2016 Mar. 27 20:47 & $78.1^\circ, -17.4^\circ$ & 11.0 \\
     & Normal & 1915 & 2016 Mar. 28 00:36 & $78.2^\circ, -17.9^\circ$ & 11.0 \\
     &        & 1921 & 2016 Mar. 27 22:42 & $78.1^\circ, -17.7^\circ$ & 11.0 \\
     &        & 1926 & 2016 Mar. 27 15:09 & $78.0^\circ, -17.1^\circ$ & 11.0 \\
\hline
Annual & Low-activity & - & $\lambda_\odot=10^\circ$ & $77^\circ, -16^\circ$ & 10.9 \\
       & Normal & - & $\lambda_\odot=10^\circ$ & $77^\circ, -16^\circ$ & 10.9 \\
\hline
\end{tabular}
\label{tbl:met}
\end{table*}

\end{CJK*}
\end{document}